\documentclass[conference]{IEEEtran}
\IEEEoverridecommandlockouts
\usepackage{cite}
\usepackage{amsmath,amssymb,amsfonts}
\usepackage{algorithmic}
\usepackage{graphicx}
\usepackage{textcomp}
\usepackage{xcolor}
\def\BibTeX{{\rm B\kern-.05em{\sc i\kern-.025em b}\kern-.08em
    T\kern-.1667em\lower.7ex\hbox{E}\kern-.125emX}}
    
\usepackage{array}
\newcolumntype{L}[1]{>{\raggedright\let\newline\\\arraybackslash\hspace{0pt}}m{#1}}
\newcolumntype{C}[1]{>{\centering\let\newline\\\arraybackslash\hspace{0pt}}m{#1}}
\newcolumntype{R}[1]{>{\raggedleft\let\newline\\\arraybackslash\hspace{0pt}}m{#1}}

\makeatletter
 \let\old@ps@headings\ps@headings
 \let\old@ps@IEEEtitlepagestyle\ps@IEEEtitlepagestyle
 \def\confheader#1{%
 \def\ps@IEEEtitlepagestyle{%
 \old@ps@IEEEtitlepagestyle%
 \def\@oddhead{\strut\hfill#1\hfill\strut}%
 \def\@evenhead{\strut\hfill#1\hfill\strut}%
 }%
 \ps@headings%
 }
\makeatother

\confheader{%
\parbox{18cm}{\centering Accepted copy for Publication at the 16th IEEE International Conference on RFID, 2022 \\ 
Final published version available at: https://doi.org/10.1109/RFID54732.2022.9795979}}

\begin{document}

\title{A Novel Secure NFC-based Approach for BMS Monitoring and Diagnostic Readout}

\author{
\IEEEauthorblockN{Fikret Basic, Claudia Rosina Laube, Christian Steger}
\IEEEauthorblockA{\textit{Institute of Technical Informatics} \\
\textit{Graz University of Technology}\\
Graz, Austria \\
\{basic, laube, steger\}@tugraz.at}
\and
\IEEEauthorblockN{Robert Kofler}
\IEEEauthorblockA{\textit{R\&D Battery Management Systems} \\
\textit{NXP Semiconductors Austria GmbH Co \& KG}\\
Gratkorn, Austria \\
robert.kofler@nxp.com}
}

\maketitle

\begin{abstract}
In modern systems that rely on the use of Battery Management Systems (BMS), longevity and the re-use of battery packs have always been important topics of discussion. These battery packs would be stored inside warehouses where they would need to be properly monitored and configured before their re-integration into the new systems. Traditional use of wired connections can be very cumbersome, and sometimes even impossible, due to the outer layers and packaging. To circumvent these issues, we propose an extension to the conventional BMS design that incorporates the use of Near Field Communication (NFC) for the purpose of wireless battery pack status readout. Additionally, to ensure that these packs are only managed by authenticated devices and that the data that is communicated with is protected against outside eavesdropping and tampering, we present a solution in the form of a lightweight security layer on top of the NFC protocol. To show the feasibility of our design, an accompanying prototype has been implemented and evaluated.
\end{abstract}

\begin{IEEEkeywords}
Battery Management System; Security; Cyber-physical; Authentication; Near field Communication; Mobile.
\end{IEEEkeywords}

\section{Introduction}
\label{sec:intro}
With the rise of general awareness for green sustainability and environmental protection, Electric Vehicles (EV) are becoming ever more prevalent. The most valued components that they contain are the battery packs. These packs lose their power over time, with many manufacturers suggesting that the battery cell packs should be replaced when the battery capacity drops to around 70\% - 80\% of their maximum capacity \cite{6228326}. While it varies, these values are expected to be reached after just ten years of active usage. To reduce the load on the living environment, reusable battery packs are almost certainly going to become important in the upcoming market, as they can be recycled for other purposes, such as for energy harvesting, or for systems with moderated safety requirements \cite{secondBatt}. 

A battery pack usually contains a set of battery cells and sensors connected to a Battery Cell Controller (BCC). The safety control and charging handling of battery packs are further managed through Battery Management Systems (BMS)~\cite{6532486}. These are specialized devices that handle the main data processing and system control from one or several battery packs, connected in a central, modular, or distributed topology \cite{10.1007/978-3-030-52794-5_13}. BMS components are traditionally coupled in an enclosed environment, and hence, work as a closed system. Therefore, when a battery pack is withdrawn to a storehouse, an external communication interface would need to be provided for the purpose of obtaining diagnostic information. The usability of this external readout is generally seen under two potential use cases: (i) warehouse-stored battery cells with their respective BCC, and (ii) active usage in systems (e.g. EVs) for faults and communication breakouts analysis. In both cases, it is of importance that the abnormal behaviour of battery cells is detected early by the BMS. Changes in temperature and storage conditions can affect the life of a battery cell \cite{HASAN2021102940, fireBattery}. Outside of the battery status readout, external communication can also be used for firmware and configuration updates \cite{7945585}.

Extending the functionality also extends the portfolio of potential malicious attacks. A capable attacker could fake a single temperature value to initiate a fake thermal runaway in the BMS. Further manipulations could even allow the attackers to completely mask the real damage that is done to the battery pack or leave an exploit that could be used to hide a fake malicious battery pack by replicating the behaviour of a real one.  It is therefore important that the communicating devices are mutually authenticated and their data adequately protected.

The readout of battery packs can be achieved by using a conventional wired interface, e.g., Controller Area Network (CAN), or other serial interfaces. However, these come with several limitations as the device handling would need to be done on an individual basis. Wireless technologies allow for more efficient handling of a larger number of battery packs. It also circumvents the limitations of packaged battery packs and allows for an external readout, with its integration into an automated environment also allowing for faster processing by employing contactless readers and assembly lines. But even with these advantages, choosing an appropriate wireless technology under the presented conditions is difficult, as we see several requirements that need to be fulfilled:
\begin{itemize}
    \item \textit{Widespread availability}: the protocol needs to be supported across multiple devices and be simple to integrate.
    \item \textit{``Wake-up'' functionality}: to reduce the reliance on the use of the battery cells and additional connection points, it is desirable that the control units function independently and are powered up from an external interface.
    \item \textit{Security considerations}: the system needs to be secure against common threats and to support an integration of the extended security communication.
\end{itemize}

%While several applications exist on the market that allow for BMS status monitoring using Bluetooth technology
To fulfil the mentioned criteria, we have decided to use Near Field Communication (NFC) as the proposed wireless technology. NFC allows for easy integration into the existing BMS architectures, offers a wide range of supported NFC readers (incl. mobile phones), and has a fast readout process. To reduce the reliance on battery cells, NFC also offers the energy harvesting feature, being able to power up an NFC-tag device from an outside NFC reader. While the NFC protocol itself does not offer a full security suite, it does offer some security features that are of an advantage when compared to other wireless technologies \cite{Haselsteiner06securityin}. Furthermore, the communication usually has smaller latency and less interference when compared to other wireless technologies used with BMS \cite{9080766}. We extend on the notion of the protocol security designs, by proposing a low-overhead security solution that can be used under the specified industrial application settings.\\
\textbf{Contributions.} Summarized, the main contributions of this paper include: (i) a design proposal for establishing external NFC readout between a configuration reader and a BMS and its battery packs, (ii) a lightweight security solution built on top of the NFC layer that is able to provide mutual authentication and secure session establishment, (iii) testing and evaluation of the presented methods on a real hardware test-suite. To the authors' best knowledge, this is the first publication that describes an NFC-design proposal with an integrated security protocol for a BMS status readout.

\section{Background and Related Work}
\subsection{Wireless Battery Management System (BMS)}
Recently, wireless BMS have become a topic of discussion since replacing wired with wireless interfaces would help in reducing production cost and complexity. Many of the recent works look for solutions using the 2.4 GHz frequency band technologies such as Bluetooth~\cite{7151581, 7357002}, ZigBee~\cite{Rahman_2017}, and WiFi~\cite{8591253}. However, most of these publications primarily focus on the inter-communication between modular BMS components, and only partially on the requirements derived for external access, which we investigate in this work.

Combining NFC applications with BMS is a relatively novel topic, as not much work has yet been done by the research community as mentioned in a current survey study of wireless BMS~\cite{electronics10182193}. A recent paper published by Basic et al.~\cite{basic_nfc} proposes a solution for wireless sensor readouts from battery cells to BCCs by using NFC technology, and also presents an anti-counterfeiting authentication measure, but only for closed active systems. In this work, we further try to bridge the gap of some of the open questions in respect to design requirements between NFC and BMS by also extending the security application for external communication interactions.

\subsection{Near Field Communication (NFC) Security}
NFC is a high-frequency Radio Frequency Identification (RFID) wireless technology operating in the 13.56MHz frequency band with a range up to 10 cm.
NFC-based tags and smart cards are typically compliant with ISO/IEC 14443 or ISO/IEC 15693.
The passive tags are able to be powered by the active readers for the duration of the data exchange. 
NDEF record is a widely accepted approach for data encapsulation in NFC, as it provides a relatively low message overhead. NFC relies on different security approaches to provide additional protection for data handling. A common approach would be to use Signature Record Type Definition (RTD). The original 1.0 version was proven to be vulnerable to attacks~\cite{5741303, 6148458}, with the 2.0 version being the one that is often deployed instead. It uses signatures through certificate chaining to provide data authenticity and integrity. However, it does not provide data confidentiality. Additionally, the employed schema relies on asymmetric cryptography, which can prove to be demanding on constrained devices, requiring a dedicated infrastructure. 

Extended solutions, like the QSNFC proposed by Ulz et al.~\cite{8376190}, provide a full security suite. QSNFC uses Diffie-Hellman key exchange and certificates for device authentication. However, this approach would not be suitable for the presented BMS use-case, as only the QSNFC's server authenticity is checked, but not the client's, leaving the possibility for configuration updates from unauthorized readers towards the BMS MCU. It also relies on  128-bit public keys, which is less than the current NIST recommendation for legacy applications. Regarding other certificate-based approaches, Urien and Piramuthu~\cite{6548149} propose a TLS schema adaptation for NFC. It, however, would be very resource-demanding for the current BMS applications and therefore not applicable.

\section{Design of a Novel Secure BMS NFC Readout}
\label{sec:design}
For the configuration and status readout of BMS, a system architecture is proposed containing the following components:
\begin{enumerate}
    \item \textit{Processing unit}: e.g. an MCU for process handling, attached either through BCC or the main BMS controller.
    \item \textit{NFC-Tag (NTAG)}: for communication and data transfer.
    \item \textit{Secure Module (SM)}: provides security functionality.
    \item \textit{Mobile reader}: a mobile device or a different NFC reader-equipped device that is also capable of the necessary processing and security operations.
\end{enumerate}
% (i) \textit{Processing unit}: e.g. an MCU for process handling, attached either through BCC or a BMS, (ii) \textit{NFC-Tag (NTAG)}: for communication and data transfer, (iii) \textit{Secure Module (SM)}: to handle security functions and key storage, (iv) \textit{External reader}: a mobile device or a different NFC reader-equipped device that is also capable of the necessary processing and security operations.

The mobile reader needs to be appropriately configured to be able to communicate with the dedicated NTAGs and BMS hardware. For the context of this work, an assumption is made which entails that the devices have been correctly pre-configured and embedded with the correct security material. %It is further assumed that the external readers are sufficiently protected from outside attacks that can happen during update sessions.
NTAG, which is used to transmit information between the BMS processing unit and the external mobile reader, is primarily used as a bridge device to pass and handle the data. This is done with the intention that the security functions and the secure data would be stored inside a trusted environment, which in this case would be an SM that resides on the battery pack together with the MCU. The NTAG can also be boosted with additional device authorization mechanisms \cite{basic_nfc}.

\begin{figure}[!ht]
  \centering
  \includegraphics[width=0.95\linewidth]{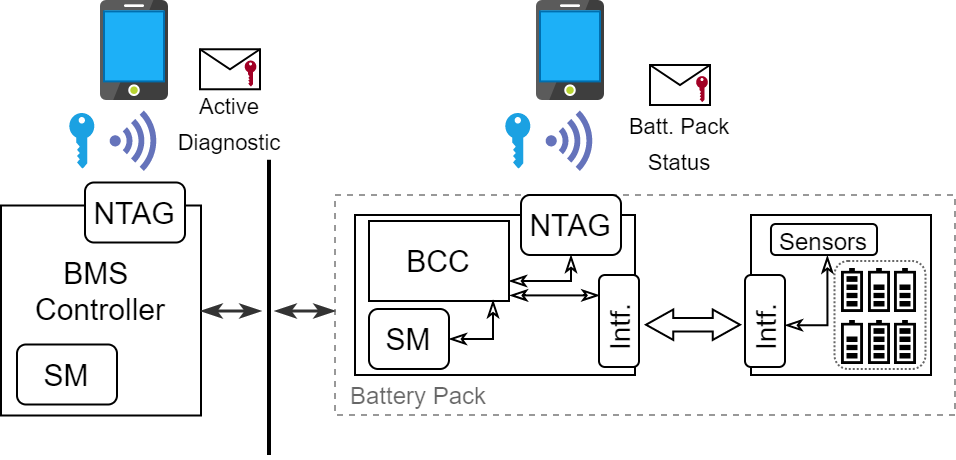}
  \caption{Proposed system architecture: The Active scenario is shown on the left-hand side with the communication going through the main BMS controller; On the right-hand side the On-Rest scenario is shown for a passive readout.}
  \label{fig:mob_readout_arch}
\end{figure}

As mentioned in Section~\ref{sec:intro}, we have focused on deriving a design solution for two specific use-case scenarios:
\begin{itemize}
    \item \textit{Active scenario}: active usage within a BMS system; capable of extracting current operational diagnostic data.
    \item \textit{On-Rest scenario}: for stored and inactive battery packs; capable of extracting lifetime and present status data.
\end{itemize}

The communication design in both cases does not change, as the security protocol stays the same. The main difference comes from the use of the NTAG component related to energy harvesting and the wake-up procedure. Namely, the wake-up procedure needs to be initiated for a stored battery pack to conserve the used energy. Here, the NTAG plays a part of the event trigger and energy supplier for the initial wake-up of the connected MCU. Based on the power draw, the MCU can either be powered directly from the NTAG, or it needs to power itself up by re-diverting the energy from the connected battery cells. As mentioned, in both cases, the actual event action does not change, and it results in a read-out of the pre-defined information. The system architecture and its applicable use cases are shown in Fig.~\ref{fig:mob_readout_arch}. The line indicates separation since for the on-rest warehouse scenario, battery packs are usually detached from their main BMS controllers. For the rest of this work, we will refer to the processing unit as the BMS for both use cases.

\subsection{Security Threats and Prerequisites}
\label{sec:sec_threats}

The communication design needs to adhere to security requirements drawn from research concerning common threats in BMS~\cite{Cheah2019}, and otherwise similar industrial systems~\cite{7005129}, as well as Confidentiality, Integrity, and Availability (CIA) principles from a general security design. Specifically, the design needs to be able to protect transferable sensitive BMS data from being spied on or tampered with. Other attacks can include a variety of replay or Man-in-the-Middle (MitM) attacks, denial of service, and malicious actions against the hardware and software integrity of a BMS device~\cite{Sripad2017,8813669, Kumbhar2018}. NFC provides a low-range communication, that limits the range from where attacks can be conducted. However, there still exist a variety of possible remote attacks that could take advantage of an unprotected channel \cite{7005129, 6861328}. These can range from sniffing attacks that can compromise the confidentiality of transferred data using eavesdropping equipment with a range of up to 10m as demonstrated by Haselsteiner and Breitfuß \cite{Haselsteiner06securityin}, up to attacks that directly target the authentication identity \cite{6975506}. With these threats in mind, authentication will also need to be provided via a mutual authentication procedure that takes place before the data exchange starts. 

\begin{figure}[!ht]
  \centering
  \includegraphics[width=0.82\linewidth]{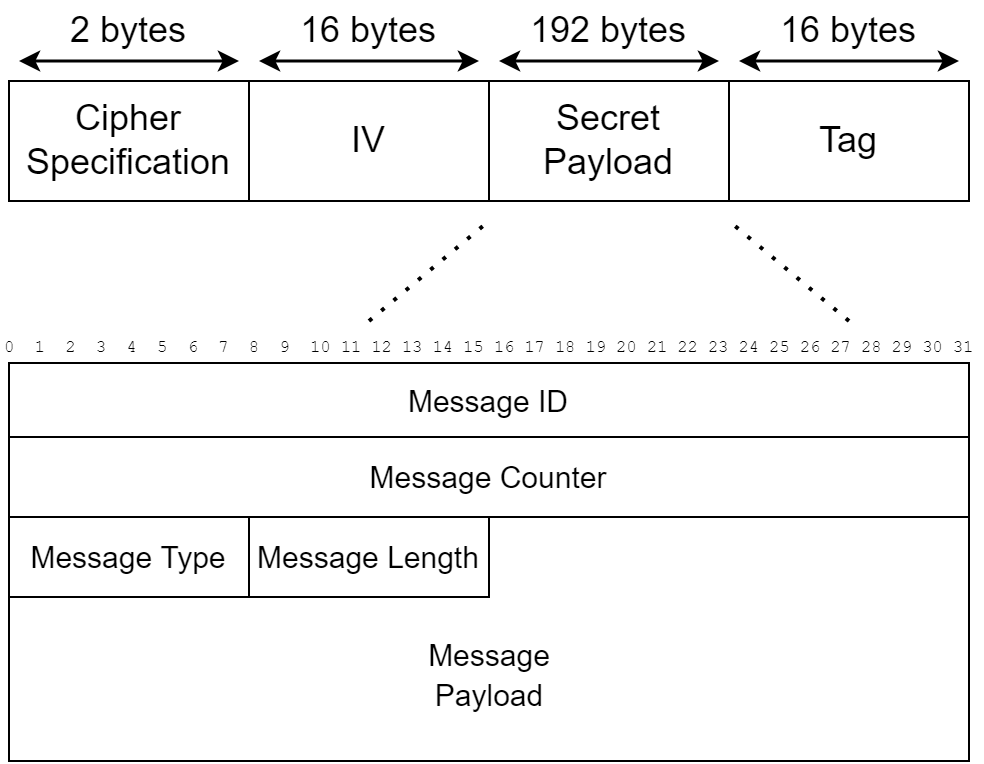}
  \caption{SNDEF record structure.}
  \label{fig:sndef_struct}
\end{figure}

An additional aim of the security design is to keep the overall structure lightweight, both in its implementation complexity, processing time, and extra data size. To achieve these properties, we opted to use a symmetric cryptography architecture approach. Security needs to be guaranteed based on Kerckhoff's principle, i.e., the master key material needs to be unique and securely stored on the devices. 

\subsection{Secure Near Field Communication Structure}

To enable the secure message exchange, a message structure has been proposed in the form of an NDEF record named Secure-NDEF (SNDEF). These records are intended to be short NDEF records, built as an extension to the proposed records from Ulz et al.~\cite{7945585}, but adapted to be more flexible in use among different cipher protocols, such as the Authenticated Encryption with Associated Data (AEAD) schemes or the traditional AES+MAC protocols. The record structure can be seen in Fig.~\ref{fig:sndef_struct}. It consists of: (i) a cipher specification (e.g., AES-CBC+CMAC, AES-GCM, AES-CCM), (ii) an Initialization Vector (IV), (iii) a secret payload, which is the encrypted data, and (iv) a tag, a piece of additional information for integrity check, e.g., a Message Authentication Code (MAC).  

The computations are done in the Encrypt-then-MAC approach, meaning that the data is first encrypted and then the tag is calculated on the data, i.e., including the secret payload and the IV. The secret payload contains a 4-bytes message ID, which for application purposes can also be adapted to be a, e.g., sensor ID. Message type holds the purpose of the action, such as \textsc{READ\_STATUS} or \textsc{UPDATE\_CONFIG}.
The structure uses the message `counter' field to keep track of a larger chain of messages. It needs to be unique for each message in a communication session (for each key) as a guard against replay attacks from rogue messages. The current design allows up message length of up to 182 bytes, which is sufficient for the application's needs, with possible extensions.

\subsection{Security Measures}

\begin{figure}[!ht]
  \centering
  \includegraphics[width=0.93\linewidth]{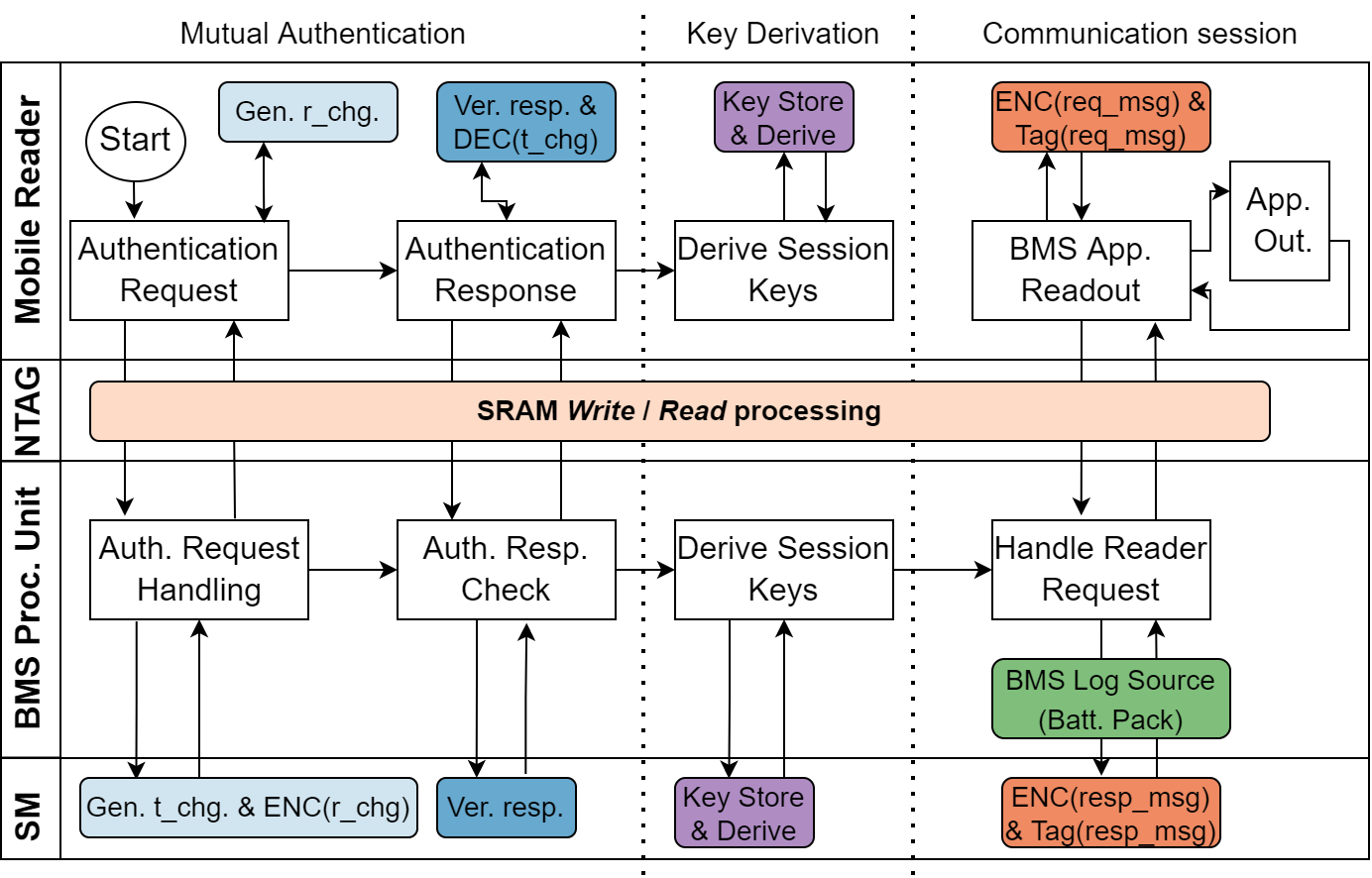}
  \caption{Diagram showcasing sequence steps between the participating devices.}
  \label{fig:main_desig}
\end{figure}

Based on the security and system analysis from Section~\ref{sec:sec_threats}, a security architecture is presented consisting of the following security protocols and operations, as seen in Fig.~\ref{fig:main_desig}.

\textbf{Mutual authentication}. Before the communication starts, both the mobile reader and the BMS module need to authenticate each other, i.e., mutually prove that they come from valid sources. The architecture uses a symmetric challenge/response mechanism with pre-embedded keys. The protocol starts with the mobile reader sending a 128-bit randomly generated challenge, to which the BMS replies with its 128-bit challenge and encryption of the reader's challenge. The reader verifies the received data and responds with the decrypted tag challenge. The BMS verifies the reader's response. 

\textbf{Keys derivation}. After the authentication, a secure channel is established. First, session keys need to be generated and derived using a Key Derivation Function (KDF) with $K_d = KDF(K_{M} \,||\, dev\_add\_data \,||\, seed \,||\, padding)$, where $K_{M}$ is the stored master key, $dev\_add\_data$ is optional and can be production data, the $seed$ is made from concatenating the nonces from the authentication step, with $padding$ being used for rounding up. 
In the case of an AEAD schema, only one key is necessary. Otherwise, a MAC key different from the encryption key is also derived from $K_{d} = (K_{Enc} \,||\, K_{MAC})$. An important detail of the design is to provide enough entropy between the authentication and key derivation procedure so as not to allow the attackers to exploit it through a replay attack. Since the authentication uses a symmetric encryption operation, the KDF function should not have blocks in its derivation that use the same keys and procedures. An example would be the Cipher-MAC (CMAC) which if used for the KDF, would also use the encryption operation possibly based on the same original key. To circumvent this, it is advisable to do either one of the following: (i) adding a guard against specialized particular nonces, i.e., not allowing re-usable challenge nonces, all zeros, etc., and using double encryption operations during the authentication step to hide the single encryption values, or (ii) to use a KDF with completely separate operations from the authentication step such as Hash-MAC (HMAC). 

\begin{figure}[!ht]
  \centering
  \includegraphics[width=0.70\linewidth]{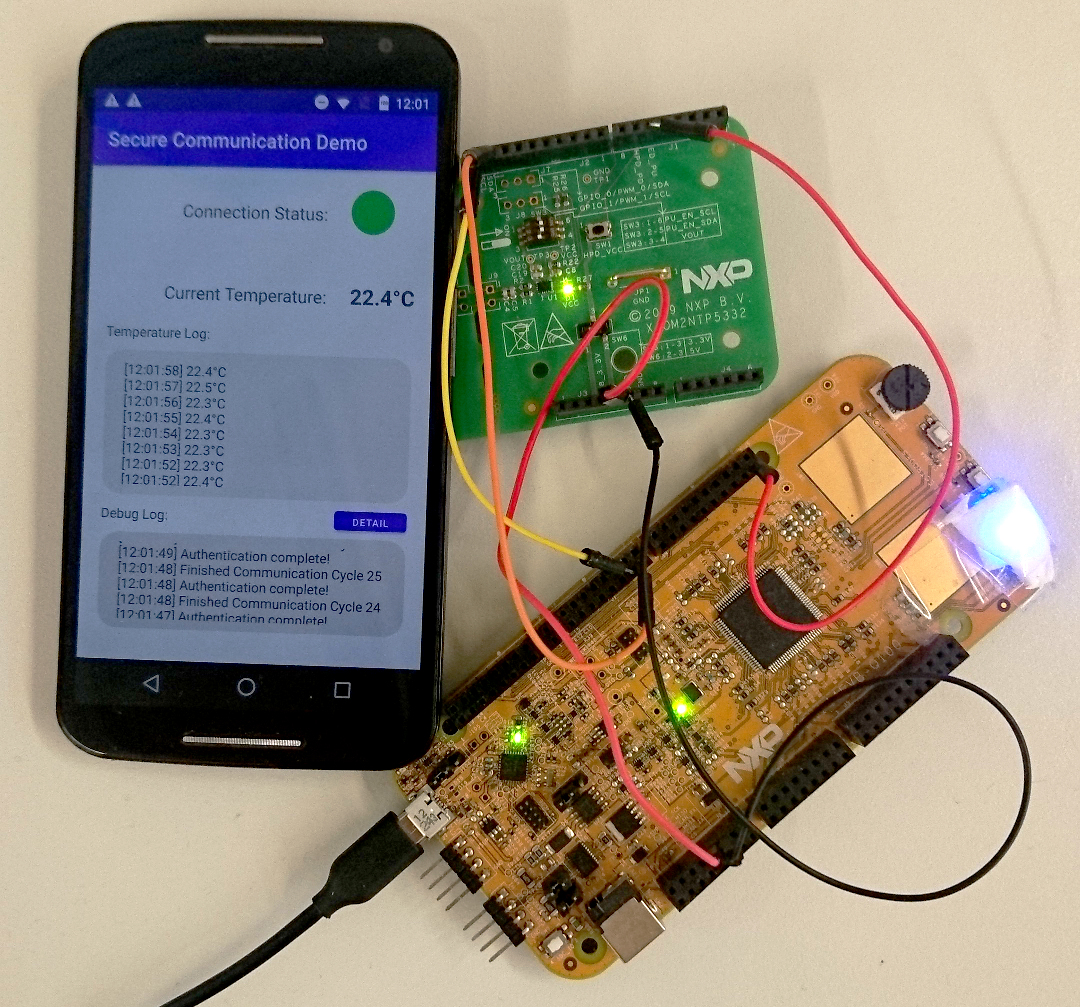}
  \caption{Prototype showcasing the secure NFC BMS readout.}
  \label{fig:prototype}
\end{figure}

\textbf{Communication session}. The mobile reader and the BMS module communicate over the NFC using the SNDEF structure. The security is provided through the use of encryption, data integrity, and authentication checks. The underlying protocols are either AES+MAC or the AEAD algorithms based on the system's availability. The Encrypt-then-MAC method is used for the best security mode protection.
The design also uses native security mechanisms found within NFC devices. This includes the limitations of the \textit{Write} command to the BMS device in case only the \textit{Read} process has been called.

\section{Evaluation}
\subsection{Prototype Implementation}
To evaluate the proposed architecture and test its feasibility in an applicable scenario, an adequate prototype was implemented. It consists of a mobile phone with an integrated NFC functionality that fulfils the role of a mobile reader. For this purpose, a Motorola Moto X running Android 6.0 was used. The BMS setup consists of NXP Semiconductor components that mimic real-world usage. An S32K144 MCU was used as the main BMS controller. It communicates with a battery pack consisting of MC33771C as the BCC and a battery cell emulator module. An NTAG Type 5 was used for the NFC interface of the BMS as an NFC-enhanced module communicating via an I2C connection.

\begin{figure}[!ht]
  \centering
  \includegraphics[width=0.75\linewidth]{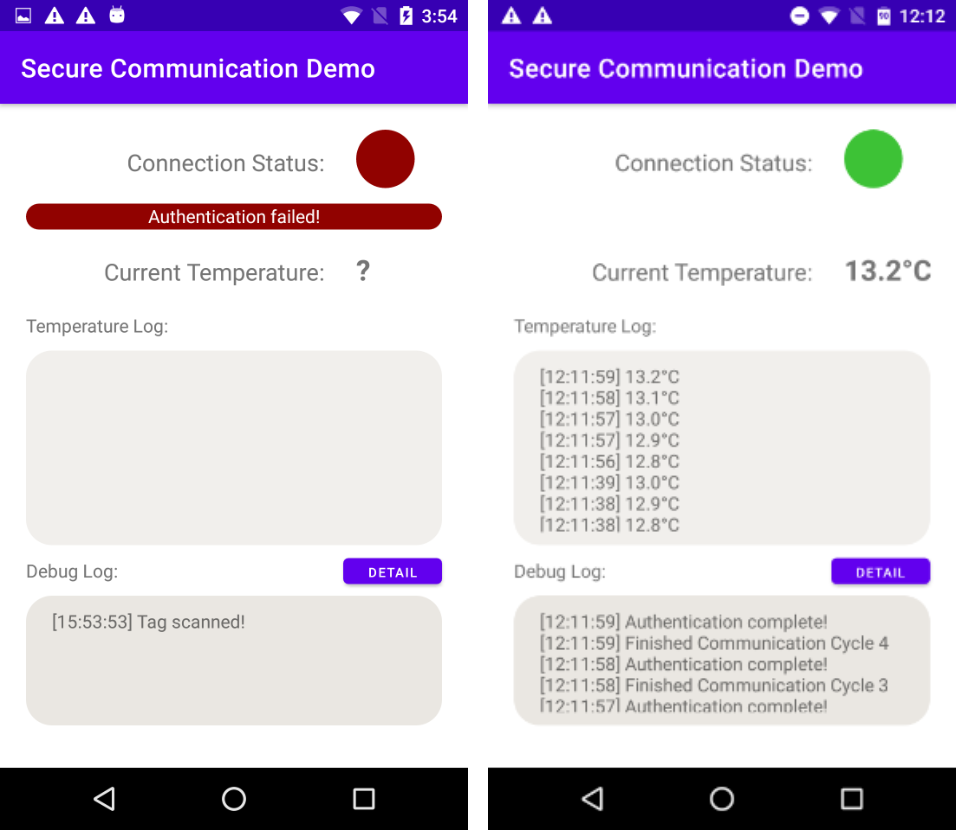}
  \caption{Developed prototype mobile application used for the evaluation.}
  \label{fig:mob_screens}
\end{figure}

The security capabilities are provided through the native Android SDK for the mobile phone, while the BMS MCU relies on an integrated Cryptographic Services Engine compressed (CSEc) \cite{csec} which implements the Secure Hardware Extension (SHE) specification \cite{autosar}. It provides basic security functions such as the Random Number Generator (RNG), secure keys storage, and AES-CBC+CMAC cipher suite, while the testing of the AE functions was done using the BearSSL security library~\cite{bearssl}. The main prototype components can be seen in Fig~\ref{fig:prototype}. Appropriate software extensions were implemented into the BMS monitoring and diagnostic firmware to handle the added protocol extensions while still allowing for the normal workflow of the basis system. Furthermore, a graphical application was developed for the mobile phone to test the usability of the main functions. Some of its application outputs can be seen in Fig~\ref{fig:mob_screens}, displaying the results after a failed and a successful authentication procedure. This setup was used for further security and performance analysis. 

\begin{figure}[!ht]
  \centering
  \includegraphics[width=0.95\linewidth]{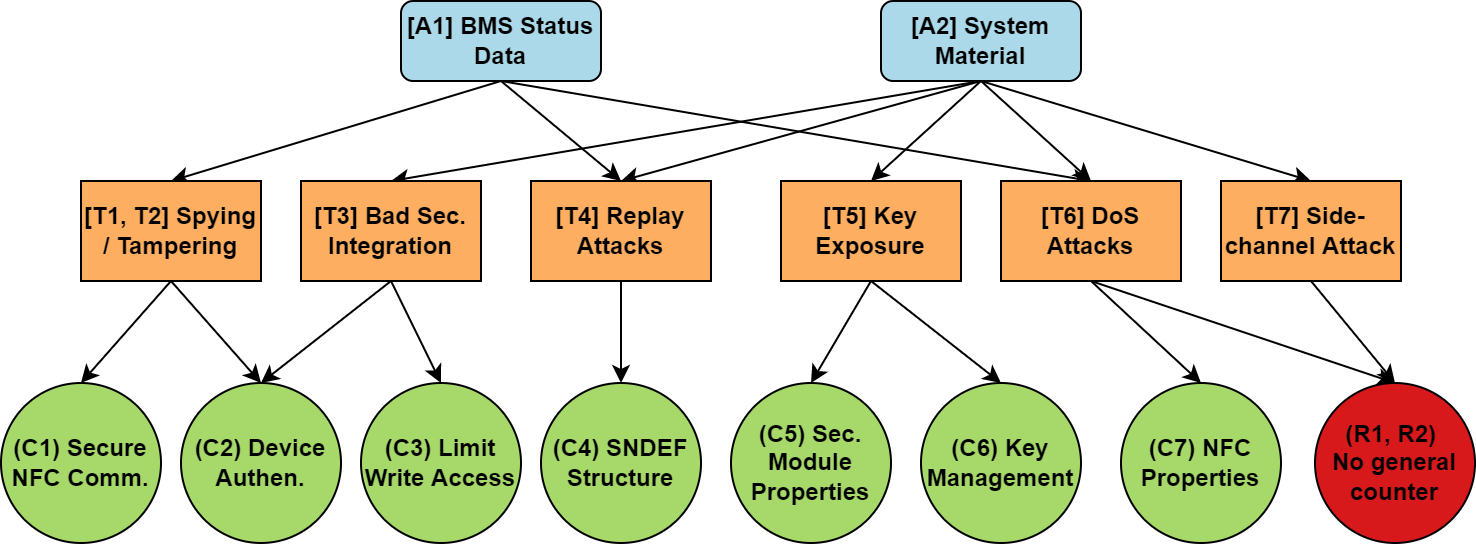}
  \caption{GSN visual model representation of the Threat analysis.}
  \label{fig:gsn_model}
\end{figure}

The conducted evaluation is applicable to both scenarios discussed in Section~\ref{sec:design} since the used security protocols remain the same.

\subsection{Security Threat Analysis}
A comprehensive security investigation has been conducted to evaluate the applicability of the proposed design \cite{Myagmar2005}. The analysis has been summarized through the specification of Assets (A), Threats (T), Countermeasures (C), and Residual Risks (R), which are derived based on the specifics of our design and investigated BMS threat concerns \cite{Cheah2019, Sripad2017, 8813669, Kumbhar2018}. An illustrative representation of the threat analysis was done using Goal Structuring Notation (GSN) modelling shown in Fig.~\ref{fig:gsn_model}. 

The system assets that need to be protected are:
\begin{itemize}
    \item $[$A1$]$ \textit{BMS status data}: functional data, i.e., diagnostic or sensor-measured data.
    \item $[$A2$]$ \textit{System configuration material}: considers general configuration data, firmware, and security material.
\end{itemize}

Each potential threat is listed followed with a short description of countermeasures, or possible residual risks, i.e., in case the threat cannot be mitigated, along with their target assets.

\begin{figure}[!ht]
  \centering
  \includegraphics[width=0.85\linewidth]{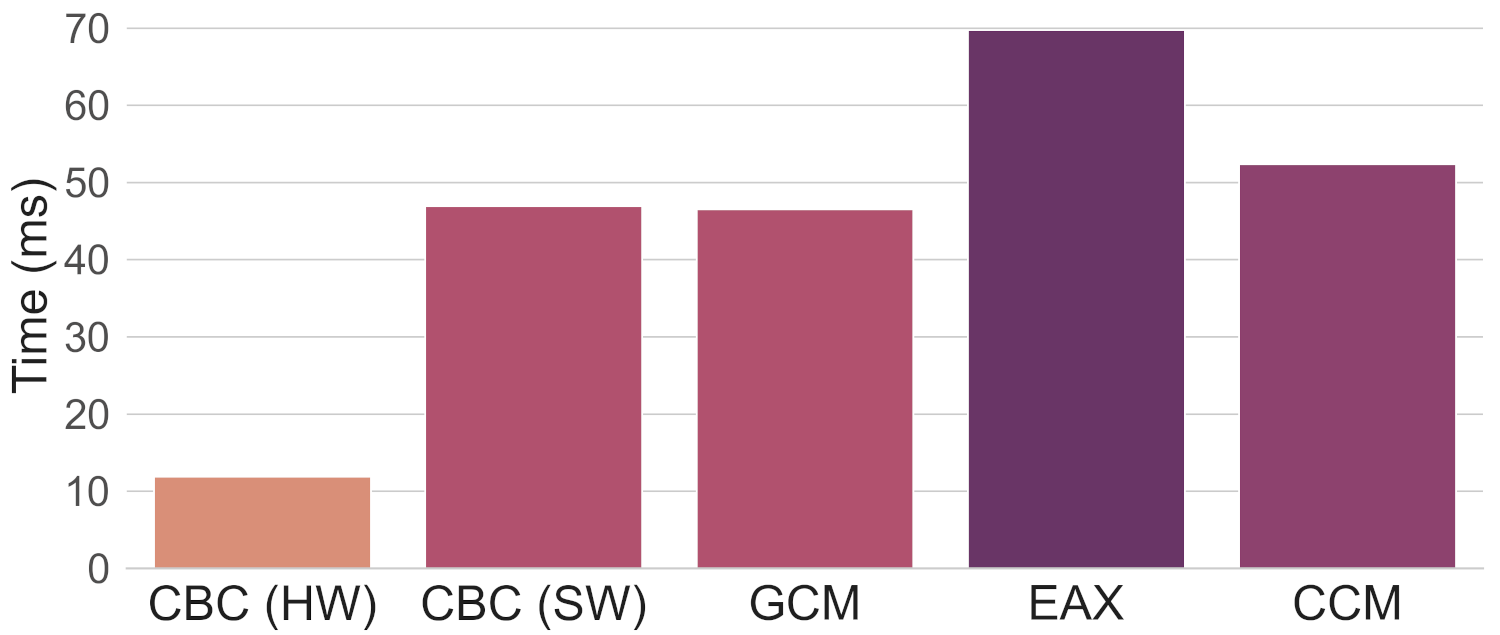}
  \caption{Crypto-algorithm operations on 192-bytes of secret data.}
  \label{fig:crypto_time_meas}
\end{figure}

\begin{itemize}
    \item $[$T1$]$ Eavesdropping on the RF channel.\\$\rightarrow (A1)$. (C1) Securing the RF channel via the proposed design using encrypted session channel with MAC check.
    \item $[$T2$]$ Channel data tampering and malicious configs.\\$\rightarrow (A1)$. (C2) Authenticating the involved parties, but also employing (C1) with MAC validation.
    \item $[$T3$]$ Faulty crypto. software implementation and bugs.\\$\rightarrow (A2)$. (C2) Entity authentication, and (C3) limiting write access to only allowed memory space.
    \item $[$T4$]$ Replay attacks through MitM manipulation.\\ $\rightarrow (A1, A2)$. (C4) SNDEF counter message field in combination with the unique key for each session.
    \item $[$T5$]$ Security material, (master, session) keys exposure.\\$\rightarrow (A2)$. (C5) Security module storage properties, (C6) key management which involves KDF \& key exchange.
    \item $[$T6$]$ Denial of Service (DoS).\\$\rightarrow (A1, A2)$. Certain attacks are partially mitigated via the (C7) NFC properties, with (R1) no general counteract. 
    \item $[$T7$]$ Side-channel attacks concerning extra ports.\\$\rightarrow (A2)$. (R2) No direct countermeasures; 
\end{itemize}

\subsection{Performance Analysis}
The first point of focus was set on comparing the performance of different AES-based encryption algorithms usable under the proposed environment. We focused on comparing the traditional AES-CBC scheme alongside AES-GCM, AES-EAX, and AES-CCM of the AEAD package. The evaluation was done on a 192-bytes application payload also including the CMAC, i.e., tag calculations. The result of the analysis can be seen in Fig.~\ref{fig:crypto_time_meas}. In addition to the software implementation of the AES-CBC, we have also compared the hardware implementation using the CSEc SM. As it can be concluded, the AES-CBC hardware execution results in the fastest time, followed by the AES-GCM for the AEAD solutions. AES-GCM generally also has better implementation support compared to other AEAD algorithms. Therefore, for the rest of the analysis, we will focus only on these two algorithms. 

\begin{figure}[!ht]
  \centering
  \includegraphics[width=0.95\linewidth]{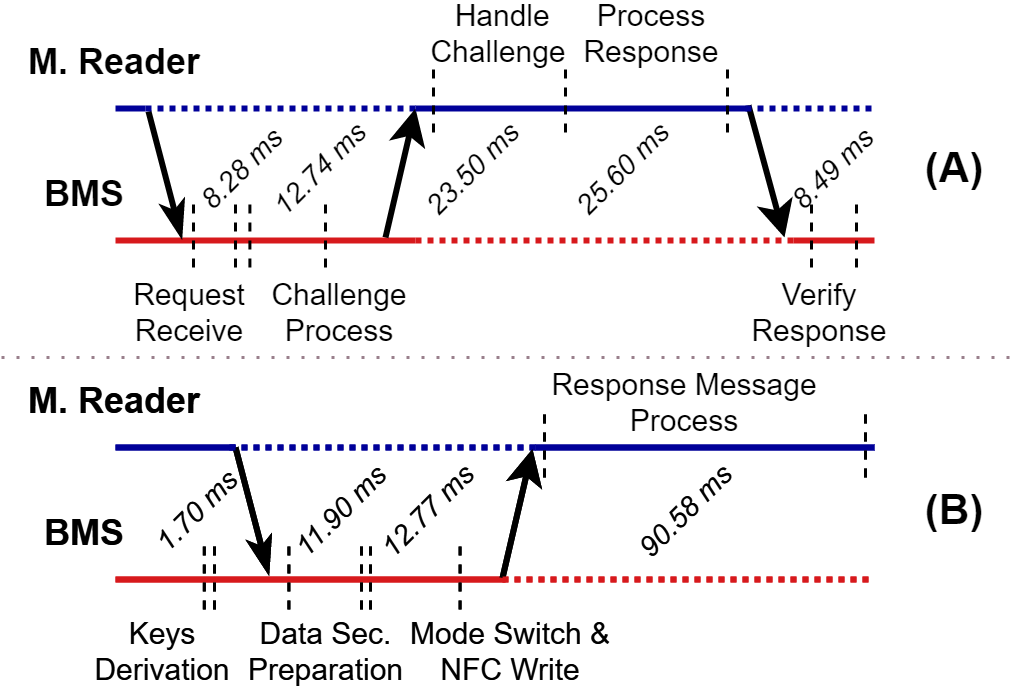}
  \caption{Timeline diagram illustrating sequence steps and presenting time measurements for: (A) Authentication phase, (B) Secure Transmission phase.}
  \label{fig:auth_phase_timeline}
\end{figure}

The performance analysis of the implemented prototype is set in a loop running enclosed process cycles. Each cycle consists of the period for the device authentication between the BMS and the mobile phone, followed by session key derivation, encryption, and exchange of 192 bytes of test data, and waiting for the readout from the mobile phone. Time measurements are derived as averages for each important cycle step after multiple runs. 
Measurements were split between the authentication and the secure transmission phase, shown in Fig.~\ref{fig:auth_phase_timeline} respectively, with the transmission step considering the AES-CBC+CMAC functions.
Each step shown considers the time of the respective data and security handling functions, as well as the NFC reading and writing operations which were the main contributors to the total execution time. 

\begin{table}[!ht] \centering
\caption{ Measured total execution times for respective process phases }
\label{table:total_exec_time}
\begin{tabular}{@{\extracolsep{0pt}} ccc}
\\[-1.8ex]\hline 
\hline \\[-1.8ex] 
\multicolumn{1}{C{1.7cm}}{Authentication} & \multicolumn{1}{C{2.4cm}}{Secure Transmission AES-CBC+CMAC} & \multicolumn{1}{C{2.4cm}}{Secure Transmission AES-GCM} \\
\hline \\[-1.8ex] 
$\:78.61\,ms\pm1.53\,ms$ & $114.34\,ms\pm1.98\,ms$ & $145.64\,ms\pm2.09\,ms$ \\
\hline \\[-1.8ex] 
\end{tabular}
\end{table}

It has been observed that repeatedly throughout the communication loop, each cycle takes roughly a constant amount of time. The time variations relied mainly on the underlying security protocols, as well as the reliability of the NFC connection, which greatly depends on the positioning of the NFC components, i.e., the position of the NFC readers relative to the NTAG. 
The total execution times are shown in Table~\ref{table:total_exec_time}. 
The resulting performance time is deemed sufficient for the intended use case.

\section{Conclusion and Future Work}
In the course of this paper, a novel system design approach was presented for a secure interaction and data exchange between a BMS and a mobile control reader. The proposed design is based on a wireless communication concept utilizing NFC technology. It is intended to be suitable for different active and passive BMS use cases, regardless of whether the data acquisition is handled through an actual BMS controller or a modulated battery pack. An NFC security record SNDEF was presented along with lightweight symmetric cryptography measures. These security enchantments provide entity authentication and a secure channel for data confidentiality and integrity protection during the mobile readout process. The SNDEF accounts both for the traditional, as well as AEAD cryptography schemes.
A demonstrative prototype was implemented for the purpose of functional verification, and security and performance evaluation. For future work, we consider an alternative design with asymmetric cryptography schemes for devices that support them. These could benefit from an expanded security architecture by offering forward secrecy and potential remote cloud support. 

\section*{Acknowledgment}
This project has received funding from the ``EFREtop: Securely Applied Machine Learning - Battery Management Systems'' (Acronym ``SEAMAL BMS'', FFG Nr. 880564).

\bibliographystyle{ieeetr}
\bibliography{references}

\end{document}